\documentclass[acmtog]{acmart}

\AtBeginDocument{%
  }

\setcopyright{acmlicensed}
\copyrightyear{2025}
\acmYear{2025}
\acmDOI{XXXXXXX.XXXXXXX}
\acmConference[SIGGRAPH '25]{Special Interest Group on Computer Graphics and Interactive Techniques Conference}
{August 10-14, 2025}{Vancouver, BC, Canada}

\usepackage{enumitem}

\begin{document}

\title{Graphics4Science: Computer Graphics for Scientific Impacts}

\author{Peter Yichen Chen}
\authornote{Corresponding Contributors
}
\email{pyc@csail.mit.edu}
\orcid{}
\affiliation{%
  \institution{University of British Columbia}  
  \city{Vancouver, BC}
  \country{Canada}}
  
\author{Minghao Guo}
\authornotemark[1]
\email{minghaog@mit.edu}
\orcid{0000-0003-3408-4997}
\affiliation{%
  \institution{CSAIL, MIT}  
  \city{Cambridge, MA}
  \country{USA}}

\author{Hanspeter Pfister}
\email{pfister@g.harvard.edu}
\orcid{}
\affiliation{%
  \institution{Harvard University}  
  \city{Cambridge, MA}
  \country{USA}}

\author{Ming Lin}
\email{lin@umd.edu}
\orcid{}
\affiliation{%
  \institution{University of Maryland College Park}  
  \city{Maryland}
  \country{USA}}

\author{William Freeman}
\email{billf@mit.edu}
\orcid{}
\affiliation{%
  \institution{CSAIL, MIT}  
  \city{Cambridge, MA}
  \country{USA}}

\author{Qixing Huang}
\email{huangqx@cs.utexas.edu}
\orcid{}
\affiliation{%
  \institution{The University of Texas at Austin}  
  \city{Austin, TX}
  \country{USA}}

\author{Han-Wei Shen}
\email{hshen@nsf.gov}
\orcid{}
\affiliation{%
  \institution{National Science Foundation}  
  \city{Alexandria}
  \country{USA}}

\author{Wojciech Matusik}
\email{wojciech@csail.mit.edu}
\orcid{0000-0003-0212-5643}
\affiliation{%
  \institution{CSAIL, MIT}  
  \city{Cambridge, MA}
  \country{USA}}

\authorsaddresses{}

\renewcommand{\shortauthors}{Chen and Guo, et al.}


\begin{abstract}
Computer graphics, often associated with films, games, and visual effects, has long been a powerful tool for addressing scientific challenges—from its origins in 3D visualization for medical imaging to its role in modern computational modeling and simulation. This course explores the deep and evolving relationship between computer graphics and science, highlighting past achievements, ongoing contributions, and open questions that remain. We show how core methods, such as geometric reasoning and physical modeling, provide inductive biases that help address challenges in both fields, especially in data-scarce settings. To that end, we aim to reframe graphics as a modeling language for science by bridging vocabulary gaps between the two communities. Designed for both newcomers and experts, \textit{Graphics4Science} invites the graphics community to engage with science, tackle high-impact problems where graphics expertise can make a difference, and contribute to the future of scientific discovery. Additional details are available on the course website: \url{https://graphics4science.github.io}.
\end{abstract}




\begin{teaserfigure}
  \centering
  \includegraphics[width=\textwidth]{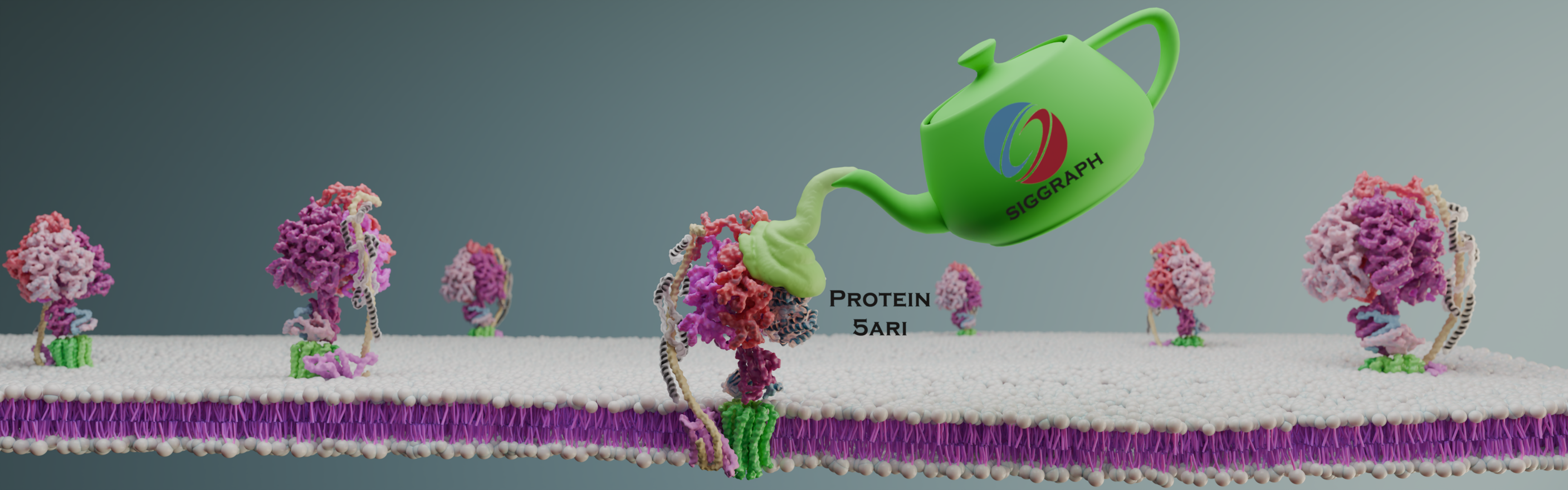}
  \Description{Teaser Figure: Graphics4Science}
  \caption{Computer graphics have evolved from a tool for visualization into a driving force behind scientific discovery, shaping advancements in biology, physics, and beyond. This course explores how graphics techniques have supported interdisciplinary research and opened up new possibilities across scientific domains.}
  \label{fig:teaser}
\end{teaserfigure}

\maketitle

\section{Introduction}

\begin{figure*}[t]
\centering
\includegraphics[width=\textwidth]{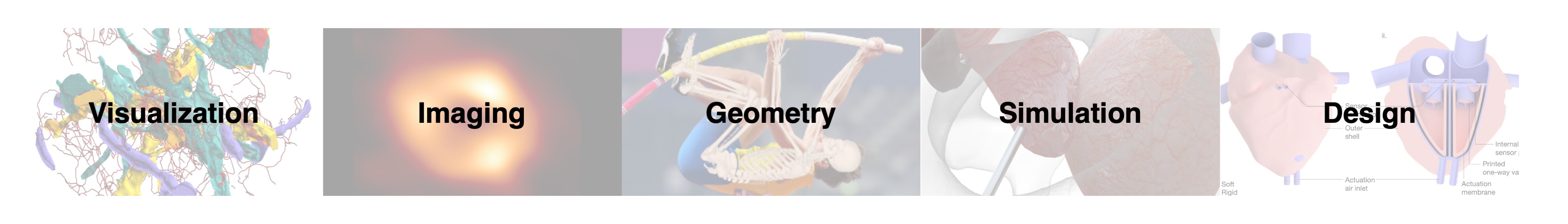}
\Description{Various computer graphics subfields for solving scientific problems.}
\captionof{figure}{\textbf{Graphics4Science.} Various computer graphics subfields play important roles in solving scientific problems. From left to right: \textbf{Visualization} enables the exploration of nanoscale astroglial cells \cite{mohammed2017abstractocyte}; \textbf{Imaging} helps unveil the first-ever image of a black hole \cite{akiyama2019first}; \textbf{Geometry} provides representations for biomechanics \cite{xia2025reconstructing}; \textbf{Simulation} allows doctors to interact with virtual human organs \cite{yang2015materialcloning}; and \textbf{Design} empowers the manufacturing of artificial hearts \cite{buchner2023vision}.}
\label{fig:combine}
\centering
\end{figure*}

Computer graphics has a strong tradition in model-based modeling. It relies on partial differential equations, rendering equations, and geometric representations to represent and simulate physical phenomena. These structures embed physical priors that serve as powerful inductive biases, supporting generalization and interpretability in computational models.

While graphics is often associated with entertainment---such as movies, games, and virtual environments---its core methods are broadly applicable to scientific problems. From medical imaging to structural biology, graphics techniques have made substantial, if sometimes underrecognized, contributions.

Both graphics and science increasingly face shared challenges: limited real-world data, high costs of acquisition, and the need for accurate simulations. In the context of machine learning, addressing these challenges requires incorporating strong inductive biases---an area where graphics’ long-standing focus on efficient, physically grounded modeling is especially valuable.

This paper introduces \textit{Graphics4Science}, an initiative aimed at foregrounding both the historical and emerging role of computer graphics in scientific discovery. Our goals are twofold: to revisit past intersections between graphics and science, and to explore new opportunities where graphics can contribute to solving pressing scientific problems.

A key barrier to deeper collaboration is the mismatch in language across disciplines. Researchers often describe similar concepts in field-specific terms---for instance, “collision” in physics versus “clashing” in protein folding. Yet these problems frequently share common computational structures: geometry, constraints, dynamics, and nonphysical states. Without a shared vocabulary, such parallels can be overlooked, and opportunities for collaboration missed.

\textit{Graphics4Science} seeks to bridge this gap. By translating scientific questions into forms familiar to the graphics community, identifying benchmarkable tasks, and showcasing examples of cross-disciplinary impact, the initiative aims to expand the scope and relevance of graphics research within science.

\section{Course Format}

\noindent\textbf{1. Introduction} (10 mins) -- Peter Yichen Chen, Minghao Guo

\smallskip
\noindent\textbf{2. Graphics for Science: Techniques \& Applications} (2 h 5 min)\\
Each talk: 20 min presentation + 5 min Q\&A
\begin{itemize}[leftmargin=8pt,labelindent=0pt,nosep]
  \item \textbf{Visualization for Science} -- Hanspeter Pfister
  \item \textbf{Simulation for Science} -- Ming Lin
  \item \textbf{Imaging for Science} -- Bill Freeman
  \item \textbf{Geometry for Science} -- Qixing Huang
  \item \textbf{Design for Science} -- Wojciech Matusik
\end{itemize}

\smallskip
\noindent\textbf{3. Panel Discussion with NSF} (45 min)\\
\textbf{Panelists:} Han-Wei Shen and selected session speakers
























\section{Description of Course}


This course aims to highlight how computer graphics techniques can extend beyond traditional entertainment applications and play a growing role in scientific research. By highlighting how graphics techniques contribute to areas such as data visualization, imaging, geometric modeling, simulation, and computational design (see Fig.~\ref{fig:combine}), the course encourages researchers and practitioners to explore new interdisciplinary opportunities. The goal is to encourage a broader view of graphics as a tool that can contribute to addressing real-world scientific questions -- from accelerating materials discovery to improving medical imaging and advancing physics-based simulations. Through this, we hope to inspire the community to bridge the gap between graphics and scientific advancements.

Rather than centering on specific technical tools, this course provides a broad vision of how graphics can drive scientific discovery. A key focus is identifying open scientific questions that the graphics community can help solve. Each session will highlight key challenges, emerging trends, and impactful applications across different scientific domains. Attendees will gain a deeper understanding of not just how graphics techniques are applied to science, but why they matter and where they could lead in the future. The course will conclude with a panel discussion, encouraging cross-disciplinary dialogue and outlining future research directions, funding opportunities, and collaborative possibilities between graphics and sciences.

\textbf{Visualization} turns unwieldy, multi-modal datasets into actionable insight. A landmark example is \emph{Foldit}~\cite{cooper2010predicting}, whose intuitive 3D interface enabled thousands of citizen scientists to manipulate protein structures and solve problems that had stumped automated algorithms. Building on such successes, this session surveys techniques that enable feature extraction and dimensionality reduction, especially for biological and medical data. We outline interactive, scalable systems that reveal hidden patterns and propose rigorous evaluation methods, showing how thoughtful visual analytics can shorten the path from data to discovery.






\textbf{Imaging} techniques that fuse physics-based forward models with data-driven priors now let scientists extract meaning from data once thought irrecoverable, from imaging the central supermassive black hole and tracking galactic dynamics~\cite{akiyama2019first}, to magnifying intracellular motion in live cells and inferring missing observations in remote-sensing streams. This session surveys super-resolution and data-synthesis methods,
comparing trade-offs in resolution, acquisition speed, and interpretability. Case studies across astronomy, microscopy, and environmental monitoring illustrate how filling in the data with informed priors accelerates discovery and opens new questions for graphics research.

\textbf{Geometry} lies at the heart of many scientific problems, providing the foundation for both representing and reasoning about physical structures. Scientific domains rely on a wide range of geometric representations---from molecular conformations and crystal lattices to protein surfaces and porous materials. In this module, we explore how geometric modeling techniques, including meshes, implicit surfaces, and neural signed distance fields (SDFs), are used to capture and manipulate complex structures across scales. We cover both synthesis (e.g., generative models for molecular structures) and analysis (e.g., structural inference from sparse data), highlighting how modern graphics techniques enable more flexible, accurate, and interpretable representations in science.

\textbf{Simulation} is a fundamental tool for modeling complex natural and engineered systems, revealing insights across spatial and temporal scales. It plays a central role in fields ranging from protein dynamics to weather prediction. To meet the demands of diverse scientific problems, a wide array of simulation techniques---such as molecular dynamics, finite element analysis, and computational fluid dynamics---have been developed. This module explores the trade-offs between accuracy, efficiency, and scalability in scientific simulations. We also examine emerging approaches, including physics-informed neural networks and differentiable solvers, that aim to accelerate computation while retaining physical fidelity. Through these perspectives, we consider how graphics-inspired simulation methods can bridge the gap between computation and real-world behavior.

\textbf{Computational Design} is transforming the way scientific problems are approached, enabling the creation of optimized materials, structures, and systems. This session will explore the role of data-driven and physics-based design in applications such as materials discovery, engineering, and generative modeling for scientific exploration. The discussion will focus on integrating simulation and optimization techniques into scientific workflows and the potential for computational design to accelerate scientific discovery.



\section{Learning Objectives and Outcomes}

By the end of this course, attendees will:  

\begin{itemize}[leftmargin=8pt,labelindent=0pt,nosep]
    \item \textbf{Recognize the role of graphics in scientific discovery} by exploring how visualization, imaging, geometry, simulation, and computational design contribute to advancements in physics, biology, materials science, and related fields.  
    \item \textbf{Examine real-world scientific applications} where graphics methods have driven breakthroughs, analyzing case studies that showcase their impact across various disciplines.  
    \item \textbf{Identify emerging research directions} in both data-driven and first-principle-informed scientific computing, highlighting key challenges and opportunities for future innovation. 
    \item \textbf{Engage with leading experts and explore collaboration opportunities} through discussions on funding strategies, industry-academia partnerships, and interdisciplinary research initiatives.  
    \item \textbf{Expand their perspective beyond entertainment applications} to understand the broader societal impact of graphics in solving critical scientific and engineering problems.  
\end{itemize}


\noindent\textbf{Looking Ahead.} This course offers only a brief glimpse into the broader landscape of graphics for science. We did not have a chance to cover many compelling directions, such as rendering for scientific discovery. We view this course as a starting point, not an endpoint, and hope to continue exploring these frontiers together through future \textit{Graphics4Science} community gatherings and collaborations.

\bibliographystyle{ACM-Reference-Format}
\bibliography{main}


\end{document}